\documentclass[12pt]{article}
\usepackage{eurosym}
\usepackage{graphicx}
\usepackage[utf8]{inputenc}
\usepackage[T1]{fontenc}
\usepackage{indentfirst}
\usepackage[margin=0.6in,nomarginpar]{geometry}
\usepackage[final]{hyperref}
\usepackage{amsmath}
\usepackage{hyperref}
\usepackage{cite}
\usepackage{xcolor}

\hypersetup{
colorlinks=true,
linkcolor=blue,
citecolor=blue,
filecolor=magenta,
urlcolor=blue
}

\begin{document}

\title{Extended uncertainty principle and Van der Waals black holes}
\author{R.Oubagha \thanks{%
rabah.oubagha@univ-oeb.dz} \\
Laboratoire de syst\`{e}mes dynamiques et contr\^{o}le (L.S.D.C), \\
D\'{e}partement des sciences de la mati\`{e}re,\\
Facult\'{e} des Sciences Exactes et SNV,\\
Universit\'{e} de Oum-El-Bouaghi, 04000, Oum El Bouaghi, Algeria. \and B.
Hamil\thanks{%
hamilbilel@gmail.com(Corresponding author)} \\
Laboratoire de Physique Math\'{e}matique et Subatomique, \\
Facult\'{e}\ des Sciences Exactes, \\
Universit\'{e}\ Constantine 1, Constantine, Algeria. \and B. C. L\"{u}tf\"{u}%
o\u{g}lu \thanks{%
bekir.lutfuoglu@uhk.cz} \\
Department of Physics, University of Hradec Kr\'{a}lov\'{e}, \\
Rokitansk\'{e}ho 62, 500 03 Hradec Kr\'{a}lov\'{e}, Czechia. \and M. Merad 
\thanks{%
meradm@gmail.com} \\
Laboratoire de syst\`{e}mes dynamiques et contr\^{o}le (L.S.D.C), \\
D\'{e}partement des sciences de la mati\`{e}re,\\
Facult\'{e} des Sciences Exactes et SNV,\\
Universit\'{e} de Oum-El-Bouaghi, 04000, Oum El Bouaghi, Algeria.}
\date{\today }
\maketitle

\begin{abstract}
In this manuscript, we investigate the extended uncertainty principle (EUP)
effects on the Van der Waals (VdW) black holes whose thermal quantities
mimic the VdW liquid. We find that the considered formalism imposes an upper
bound on the event horizon radius. Thus, the mass, Hawking temperature, and
heat capacity become physically meaningful within a certain range of event
horizon radii. At a large event horizon radius the black hole has a remnant. 
{\color{red} We observe that for a given set of parameters, the VdW black
hole can be completely unstable for all horizon radii, while for another set
of parameters, it can be unstable or stable depending on the horizon radius.}
\end{abstract}


\newpage

\section{Introduction}

An important prediction of several candidates of quantum gravity, such as
string theory \cite{amati}, loop quantum gravity \cite{roveli}, and
non-commutative geometry \cite{Girelli} is the being of a fundamental
distance at the order of the Planck length $\ell _{P}=10^{-35}\mathrm{m}$.
This length cannot be directly detected using existing technology. This is
due to the fact that the quantum gravity effects are predicted to be
directly observable only at energy levels on the order of the Planck energy $%
\sim 10^{19}$ \textrm{GeV}, a magnitude $15$ times greater than the energy
scales currently attainable by the Large Hadron Collider. This is why
indirect measurements of the Planck length are relied on to investigate
quantum gravity experimentally. On the other hand, the existence of a
minimum length in the quantum mechanical framework can also be explained by
a deformation of Heisenberg's usual uncertainty principle (HUP). The new
formalism, which is called the generalized uncertainty principle (GUP) in
the literature, is given by a modification of the position and/or momentum
operators of the Hilbert space \cite{kempf,kempf1,kempf2}. Such
modifications can also change the curvature of spacetime \cite{kempf,
kempf1, kempf2, Bolen, Park}. For example, a deformed algebra that defines
the minimum measurable momentum concept is known in the literature as the
Extended Uncertainty Principle (EUP). Moreover, in an interesting work \cite%
{S. Mignemi}, Mignemi has shown that the EUP formalism can be obtained if
quantum mechanics is defined on an (anti)-de Sitter (AdS) background with an
appropriately selected parametrization.

On the other hand, one of the most exciting topics in present times is the
thermodynamics of gravitational objects which includes the possibility of
associating concepts like temperature, pressure, volume, and entropy with
them. In the classical approach of general relativity, a black hole cannot
emit radiation. However, in 1975 through a semi-classical perspective,
Hawking demonstrated that a black hole could indeed emit radiation \cite{1}.
In the following year, he examined the radiation through the Wick Rotation
method and proposed that if the quantum effects are taken into account, then
black holes can certainly radiate \cite{2,3}. This discovery, later called
Hawking radiation, lead to significant progress in the field and convince
scientists to discuss the thermal quantities of black holes via a measurable
thermodynamic temperature. Other than Hawking's approach, several other
methods have been proposed in the literature to predict the temperature of a
black hole \cite{can1, can2, can3, can4}. One of them, which will be used in
this manuscript, is based on the surface gravity of the black hole and it
employs the zero and first laws of black hole thermodynamics. It is worth
noting that the black hole's entropy functions are assumed to be linearly
proportional to their event horizon areas in Planck units, and according to
the second law of black hole thermodynamics, their surface areas, thus
entropies, do not lessen \cite{4,5}.

The thermodynamic properties of asymptotically AdS black holes have prompted
authors to study the AdS/CFT correspondence. In \cite{6}, Hawking and Page
discussed a first-order phase transition between the thermal AdS space and
Schwarzschild AdS black hole. In an interesting study \cite{7}, the authors
showed that when they generalized the Schwarzschild AdS black hole to the
case of a charged or rotating black hole, they obtained the same behavior as
the Van der Waals (VdW) fluid. Moreover, they claimed that the analogy is
further strengthened if one considers the extended phase space instead with
a negative cosmological constant which corresponds to the thermodynamic
pressure \cite{8}. Inspired by this qualitative analogy, Rajagopal et al.
presented a black hole metric whose thermodynamics mimic the VdW fluid \cite%
{9}. In the same year, Delsate and Mann extended the metric form to $d$
dimensions by assuming a particular stress tensor form \cite{10}. They
concluded that the latter metric could be considered a near-horizon one. In
2016, Parthapratim discussed enthalpy, geometric volume and quantum
corrected-entropy functions of the VdW black hole \cite{Parthapratim}.
Later, Hu et al. handled the same system with Quevedo's
geometrothermodynamic method \cite{Huchen}. {\color{red} In a very recent
work, Luciano and Sheykhi presented an interesting discussion between black
hole geometrothermodynamics and critical phenomena within the Tsallis
entropy-based perspective \cite{Luciano}. } For further details, we refer to
the review article \cite{mangal}. Other gasses' thermal quantities were also
shown to be correlated with black hole thermodynamics. For example, in ref.%
\cite{11}, Setare and Adami introduced a Polytropic black hole, whose
spacetime is asymptotically AdS, with the same thermal properties of a
Polytropic gas. Debnath derived a novel black hole metric whose
thermodynamics matches with a special case of the Chaplygin gas model in 
\cite{12}.

Recently, \"Okc\"u and Aydiner revisited the VdW black hole thermodynamics
within the GUP formalism and discussed the quantum deformation effects by
comparing their findings with the original ones \cite{13}. Keeping in mind
the rich content of the EUP formalism, which differs from the GUP formalism,
in this paper we aimed to investigate the VdW black hole and its
thermodynamics in the EUP formalism. To this end, we constructed the
manuscript as follows: In Sect. \ref{sec2}, we briefly introduced the
considered EUP formalism and presented its effect on the Hawking temperature
with a semi-classical approach. Then, in Sect. \ref{sec3} we examined the
EUP-corrected VdW black hole thermodynamics. In Sect. \ref{sec4}, we
revisited our results by reducing them to {\color{red}three} special
sub-cases. Finally, we ended the manuscript with a brief conclusion.

\section{Extended uncertainty principle and Hawking temperature}

\label{sec2}

In this study, we consider the most basic form of the EUP formalism, which
is examined comprehensively in \cite{kempf, kempf1, kempf2, Bolen, Park, S.
Mignemi}, in AdS space: 
\begin{equation}
\Delta X\Delta P\geq \hbar \left( 1+\beta \left( \Delta X\right) ^{2}\right).
\end{equation}
Here, $\beta$ is the deformation parameter in the form of $\beta =\frac{%
\beta _{0}}{L_{\ast }}$, with a dimensionless parameter, $\beta _{0}$, and a
large fundamental distance scale, $L_{\ast}$. Undoubtedly, the most
important feature of this quantum deformation is that it sets an absolute
lower bound on the momentum uncertainty, $\left( \Delta P\right)
_{\min}=\hbar \sqrt{\beta }$. This is why it has recently attracted
increasing interest \cite{Raimundo, bilel, bilel1, bilel2, bilel3, Hassan,
Hassan1, Hassan2, Mariusz, Jaume, Ozgur, Ali, dahbi, Pet1, Mur, HH, HC}.

On the other hand, in the semi-classical approach of black hole
thermodynamics, the Hawking temperature is defined by 
\begin{equation}
T=\frac{\kappa }{8\pi }\times \frac{dA}{dS},
\end{equation}
where $\kappa $, $A$, $S$ denote the surface gravity at the outer horizon,
surface area and entropy, respectively. According to the heuristic approach,
a particle absorption can cause a minimal change in the black hole's surface
of the form \cite{Xiang}: 
\begin{equation}
\Delta A\simeq \Delta X\Delta P.
\end{equation}
Assuming that the position uncertainty on the horizon is proportional to the
event radius, $\Delta X=r_{H}$, we can express the minimal increase of the
surface area in the EUP case as 
\begin{equation}
\Delta A\simeq \frac{\gamma \hbar }{2}\left( 1+\beta r_{H}^{2}\right).
\end{equation}
Here, $\gamma$ is the calibration factor that tunes the result to the HUP
limit. As the result of particle absorption a minimal increase in black hole
entropy could be taken as $\left( \Delta S\right)_{\min }=\ln 2$, then we
have 
\begin{equation}
\frac{dA}{dS}\simeq \frac{\left( \Delta A\right) _{\min }}{\left( \Delta
S\right) _{\min }}=\frac{\gamma \hbar }{2\ln 2}\left( 1+\beta
r_{H}^{2}\right) .
\end{equation}%
After determining the calibration factor in the limit of $\beta \rightarrow
0 $ as $\gamma=4\ln 2$, we can express the EUP-corrected Hawking temperature
as 
\begin{equation}
T=\frac{\hbar\kappa }{4\pi }\left( 1+\beta r_{H}^{2}\right).  \label{for}
\end{equation}%
Hereafter, we set $\hbar=1$ for simplicity.

\section{EUP-corrected thermodynamic of Van der Waals black holes}

\label{sec3}

In the literature, the VdW equation of state describing the VdW fluid is
expressed in closed form by a two-parameter equation \cite{9}: 
\begin{equation}
T=\left( P+\frac{a}{v^{2}}\right) \left( v-b\right).  \label{eq}
\end{equation}
Here, $a$ and $b$, are two positive values that measure the attraction
between fluid molecules and their volume, and $v$ is the specific volume.

Now, we have to construct an asymptotically AdS black hole metric that
precisely mimics the thermodynamics of the considered VdW fluid equation of
state. To this end, we follow the static spherically symmetric case ansatz 
\begin{equation}
ds^{2}=-f\left( r\right) dt^{2}+\frac{1}{f\left( r\right) }%
dr^{2}+r^{2}\left( d\theta ^{2}+\sin ^{2}\theta d\phi ^{2}\right) ,
\label{L}
\end{equation}%
with the lapse function, $f\left( r\right) $, of the following form 
\begin{equation}
f\left( r\right) =\frac{r^{2}}{l^{2}}-\frac{2M}{r}-h\left( r,P\right) .
\label{f}
\end{equation}%
Here, $M$ is the black hole mass, $h\left( r,P\right) $ is an unknown
function that has to be determined, $l^{2}$ is the AdS radius, and $P$ is
the thermodynamic pressure defined by 
\begin{equation}
P=\frac{3}{8\pi l^{2}}.
\end{equation}
\textcolor{red}{We make the assumption that this metric represents a solution to the
Einstein field equations with a given energy-momentum source, 
\begin{equation}
G_{\mu \nu }+\Lambda g_{\mu \nu }=8\pi T_{\mu \nu }.  \label{G}
\end{equation}
} 
\textcolor{red}{Here, we choose the stress-energy tensor, $T_{\mu \nu }$, as an anisotropic fluid source taking the following particular form
\begin{equation}
T^{\mu \nu }=\rho e_{0}^{\mu }e_{0}^{\nu }+\sum_{i}p_{i}e_{i}^{\mu
}e_{i}^{\nu }.  \label{T}
\end{equation}Here, $\rho $ is the energy density and $p_{i}$ is principal pressures ($i=1,2,3$). Now, by employing Eqs. (\ref{L}), (\ref{G}) and (\ref{T}) we get\begin{equation}
\rho =-p_{1}=\frac{1-f-rf^{\prime }}{8\pi r^{2}}+P,
\end{equation}
and
\begin{equation}
p_{2}=p_{3}=\frac{rf^{\prime \prime }+2f^{\prime }}{16\pi r}-P.
\end{equation}
} 
\textcolor{red}{To ensure the existence of a valid physical solution, the stress-energy tensor should necessarily satisfy specific energy conditions.
\begin{eqnarray}
&&\text{Weak: } \ \ \ \ \rho \geq 0, 
 \text{ \ \ \ \ \ \ and, \ \ }\rho +p_{i}\geq 0, \\
&&\text{Strong: } \ \ \ \rho +p_{i}\geq 0, \text{  and, \ \ }\rho +\sum_{i}p_{i}\geq 0, \\
&&\text{Dominant: }\rho \geq 0, \text{ \ \ \ \ \ and, \ \ \ }\rho -\left\vert
p_{i}\right\vert \geq 0.
\end{eqnarray}
} 
\textcolor{red}{
After determining the function $h\left( r,P\right) $, hence, $f\left(r\right) $, we are going to verify these conditions mentioned.}

\begin{figure}[tbh]
\begin{minipage}[t]{0.5\textwidth}
        \centering
        \includegraphics[width=\textwidth]{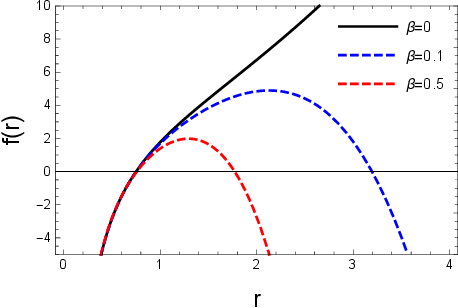}
       \centering{(a) $ a=1$, $b=3/8\pi$.}\label{figEUPLapse1a}
   \end{minipage}%
\begin{minipage}[t]{0.50\textwidth}
        \centering
        \includegraphics[width=\textwidth]{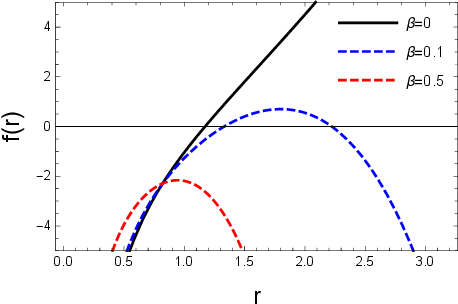}\label{figEUPLapseb}\\
        \centering{(b) $ a=1/2\pi$, $b=1$. }
    \end{minipage}\hfill
\caption{The variation of the EUP-lapse function versus radius for $M=1$ and 
$P=0.1.$}
\label{figEUPLapse}
\end{figure}

Now, using $f\left( r_{H}\right) =0$ we can express the mass in terms of the
event horizon radius, $r_{H}$ as 
\begin{equation}
M=\frac{4}{3}\pi r_{H}^{3}P-\frac{r_{H}}{2}h\left( r_{H},P\right) .
\label{M}
\end{equation}%
Next, we derive the thermodynamic volume 
\begin{equation}
V=\frac{\partial M}{\partial P}=\frac{4}{3}\pi r_{H}^{3}-\frac{r_{H}}{2}%
\frac{\partial }{\partial P}h\left( r_{H},P\right) ,  \label{V}
\end{equation}%
and then we express the specific volume using $v=6V/N$, where, $N$ stands
for the degrees of freedom%
\cite{mangal} 
\begin{equation}
v=\frac{6}{4\pi r_{H}^{2}}\left[ \frac{4}{3}\pi r_{H}^{3}-\frac{r_{H}}{2}%
\frac{\partial }{\partial P}h\left( r_{H},P\right) \right] .  \label{SV}
\end{equation}%
After that, with the help of Eqs. \eqref{for}, \eqref{f} and \eqref{M} we
derive the EUP-corrected Hawking temperature of the VdW black hole in terms
of undetermined function as 
\begin{equation}
T=\frac{1}{4\pi }\left( 8\pi r_{H}P-\frac{1}{r_{H}}h\left( r_{H},P\right) -%
\frac{\partial }{\partial r_{H}}h\left( r_{H},P\right) \right) \left(
1+\beta r_{H}^{2}\right) .  \label{Tem}
\end{equation}%
By assuming that $h\left( r,P\right) =A\left( r\right) -PB\left( r\right) $
and using the equality between (\ref{eq}) and (\ref{Tem}), we obtain an
equation of the form%
\begin{equation*}
F_{1}\left( r_{H}\right) +PF_{2}\left( r_{H}\right) =0,
\end{equation*}%
where the new defined functions, $F_{1}\left( r\right) $ and $F_{2}\left(
r\right) $, depend on the $A\left( r\right) $ and $B\left( r\right) $
functions and their derivatives. One trivial solution to this two-component
equation is that each component can be equal to zero. This can be achieved
by solving the following two ordinary differential equations simultaneously. 
\begin{eqnarray}
v-b &=&\frac{1}{4\pi }\left( 8\pi r_{H}+\frac{1}{r_{H}}B\left( r_{H}\right) +%
\frac{\partial }{\partial r_{H}}B\left( r_{H}\right) \right) \left( 1+\beta
r_{H}^{2}\right) ,  \label{B} \\
\frac{a}{v^{2}}\left( v-b\right) &=&-\frac{1}{4\pi }\left( \frac{1}{r_{H}}%
A\left( r_{H}\right) +\frac{\partial }{\partial r_{H}}A\left( r_{H}\right)
\right) \left( 1+\beta r_{H}^{2}\right) .  \label{A}
\end{eqnarray}%
By substituting Eq. (\ref{SV}) into Eq, (\ref{B}), we find 
\begin{equation}
2r_{H}+\frac{3}{4\pi r_{H}}B\left( r_{H}\right) -b=\frac{1}{4\pi }\left(
8\pi r_{H}+\frac{1}{r_{H}}B\left( r_{H}\right) +\frac{\partial }{\partial
r_{H}}B\left( r_{H}\right) \right) \left( 1+\beta r_{H}^{2}\right) .
\label{eqss}
\end{equation}%
In the first order of $\beta $, Eq. (\ref{eqss}) yields the following
solution 
\begin{equation}
B\left( r_{H}\right) =-4\pi r_{H}^{2}\left( 2b\beta r_{H}-\frac{b}{r_{H}}+%
\frac{2}{3}\right) +c_{1}r_{H}^{2}\left( 1-\frac{3}{2}\beta r_{H}^{2}\right)
,  \label{sb}
\end{equation}%
with an integration constant. To determine $c_{1}$, we can go to the limit
of $\beta \rightarrow 0$, where Eq. (\ref{sb}) should reduce to the HUP
result given in \cite{9}. In so doing, we find $c_{1}=\frac{8\pi }{3}$.
Thus, Eq. (\ref{sb}) reads 
\begin{equation}
B\left( r_{H}\right) =4\pi br_{H}-8\pi \beta r_{H}^{3}\left( b+\frac{r_{H}}{2%
}\right) .  \label{sbb}
\end{equation}%
Then, following similar straightforward algebra, we find that Eq, (\ref{sbb}%
) can be expressed as follows 
\begin{eqnarray}
\frac{\partial }{\partial r_{H}}A\left( r_{H}\right) &=&-\frac{1}{r_{H}}%
A\left( r_{H}\right) +\frac{4\pi ab}{\left( 2r_{H}+3b\right) ^{2}}\left[ 1+%
\frac{6\beta r_{H}^{2}}{2r_{H}+3b}\left( 2b+r_{H}\right) \right] +  \notag \\
&&\frac{4\pi a\beta r_{H}^{2}}{\left( 2r_{H}+3b\right) }-\frac{4\pi ab\beta
r_{H}^{2}}{\left( 2r_{H}+3b\right) ^{2}}-\frac{4\pi a}{\left(
2r_{H}+3b\right) }\left[ 1+\frac{3\beta r_{H}^{2}}{2r_{H}+3b}\left(
2b+r_{H}\right) \right] ,
\end{eqnarray}%
which leads to the following solution of the form 
\begin{equation}
A\left( r_{H}\right) =\frac{4\pi ab}{r_{H}}\log (\frac{3}{2}+\frac{r_{H}}{b}%
)+\frac{81\beta \pi ab^{5}}{8(3b+2r_{H})^{2}}+\frac{3\pi ab^{2}\left(
8-45\beta b^{2}\right) }{8r_{H}\left( 3b+2r_{H}\right) }-2\pi a\left( 1+%
\frac{\beta r_{H}^{2}}{6}-\frac{\beta br_{H}}{2}+\frac{9\beta b^{2}}{8}%
\right) .  \label{saa}
\end{equation}%
Then, we combine Eqs. (\ref{sbb}) and (\ref{saa}) and obtain the function $%
h\left( r,P\right) $ 
\begin{eqnarray}
h\left( r,P\right) &=&-2\pi a-\frac{9}{4}\beta \pi ab^{2}+\frac{4\pi ab}{r}%
\log (\frac{3}{2}+\frac{r}{b})-\frac{\beta \pi ar^{2}}{3}+\frac{81}{8}\frac{%
\beta \pi ab^{5}}{(3b+2r)^{2}}+\frac{3}{8r}\frac{\pi ab^{2}\left( 8-45\beta
b^{2}\right) }{3b+2r}  \notag \\
&&+\pi \left( \beta a-4P\right) br+8\pi \beta Pr^{3}\left( b+\frac{r}{2}%
\right) .  \label{h}
\end{eqnarray}%
which is needed for the lapse function 
\begin{eqnarray}
f\left( r\right) &=&\frac{8\pi Pr^{2}}{3}-\frac{2M}{r}+2\pi a+\frac{9}{4}%
\beta \pi ab^{2}-\frac{4\pi ab}{r}\log (\frac{3}{2}+\frac{r}{b})+\frac{\beta
\pi ar^{2}}{3}-\frac{81}{8}\frac{\beta \pi ab^{5}}{(3b+2r)^{2}}  \notag \\
&-&\frac{3}{8r}\frac{\pi ab^{2}\left( 8-45\beta b^{2}\right) }{3b+2r}-\pi
\left( \beta a-4P\right) br-8\pi \beta Pr^{3}\left( b+\frac{r}{2}\right) .
\label{METFUN}
\end{eqnarray}%
The present form of the lapse function describes the VdW black hole in the
EUP formalism up to the first order of $\beta $. It is worth noting that for 
$\beta =0$, it reduces to 
\begin{equation*}
f_{HUP}(r)=2\pi a-\frac{2M}{r}+\frac{8\pi Pr^{2}}{3}\left( 1+\frac{3b}{2r}%
\right) -\frac{4\pi ab}{r}\log (\frac{3}{2}+\frac{r}{b}),
\end{equation*}%
which is the same as the Eq. (18) in \cite{9}. Hereafter, the thermal
properties, which we will obtain, will represent the thermodynamics of the
EUP-VdW black hole. Before doing that, we depict the first-order
EUP-corrected lapse function versus radius in Fig. \ref{figEUPLapse}. 

\newpage We observe that the EUP parameter modifies the lapse function with
the second turning point. In other words, in the ordinary case, the lapse
function increases monotonically however, in the EUP case it does not,
therefore the EUP modification changes the VdW black hole geometry and its
thermodynamics. The new zero points correspond to new singularities and we
will investigate them over the thermal quantities. 
\textcolor{red}{In
addition, we note that when $r$ is smaller than one, all the cases are
covered by each other in Fig. 1(a), but we cannot see this behavior in Fig.
1(b). This is due to the weak intermolecular attraction between the fluid
constituents for $a<<1$.}

{\color{red}In Fig. \ref{enefig} we examine the energy conditions. We see
that the EUP corrections do not alter the energy conditions as found in the
non-corrected case. 
\begin{figure}[tbh]
\begin{minipage}[t]{0.33\textwidth}
        \centering
        \includegraphics[width=\textwidth]{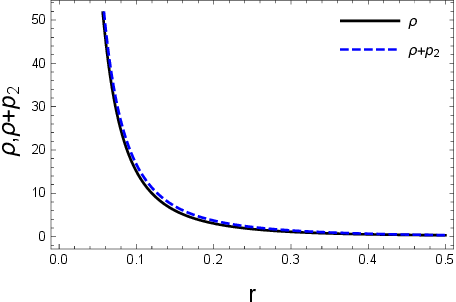}
       \centering{(a) Weak energy condition.}\label{Weakenergy}
   \end{minipage}%
\begin{minipage}[t]{0.33\textwidth}
        \centering
        \includegraphics[width=\textwidth]{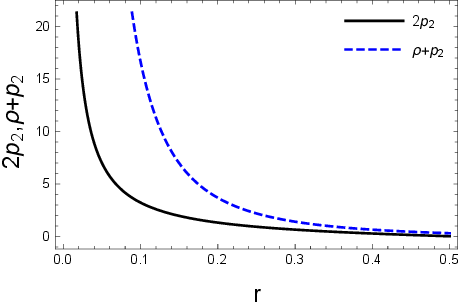}\label{Strongenergy}\\
        \centering{(b) Strong energy condition. }
    \end{minipage}
\begin{minipage}[t]{0.33\textwidth}
        \centering
        \includegraphics[width=\textwidth]{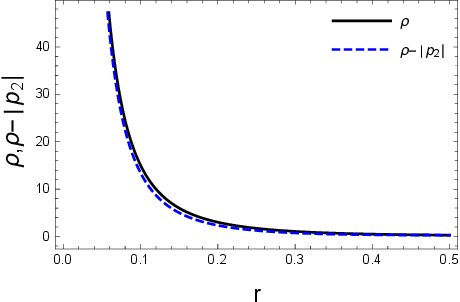}
       \centering{(c) Dominant energy condition.}\label{domenergy}
   \end{minipage}\hfill
\caption{Energy conditions are plotted for $a=\frac{3}{2\protect\pi }%
,b=1,m=0.1,P=0.1$ and $\protect\beta =0.5.$ }
\label{enefig}
\end{figure}
}

Now, let us use Eqs. \eqref{M} and \eqref{V} to re-express the EUP-corrected
VdW black hole mass 
\begin{eqnarray}
M &=&-\frac{81}{16}\frac{a\beta b^{5}}{(3b+2r_{H})}\pi r_{H}+\frac{9\pi }{8}%
a\beta b^{2}r_{H}-\frac{3\pi ab^{2}\left( 8-45\beta b^{2}\right) }{%
16(3b+2r_{H})}-\frac{\pi b}{2}r_{H}^{2}(a\beta -4P)  \notag \\
&&-2\pi ab\log \left( \frac{r_{H}}{b}+\frac{3}{2}\right) +\frac{4\pi }{3}%
Pr_{H}^{3}+\pi ar_{H}+\frac{1}{6}\pi a\beta r_{H}^{3}-4\pi \beta
Pr_{H}^{4}\left( b+\frac{r_{H}}{2}\right) ,  \label{MA}
\end{eqnarray}%
and EUP-corrected VdW black hole volume 
\begin{equation}
V=2\pi br_{H}^{2}+\frac{4\pi }{3}r_{H}^{3}-4\pi \beta r_{H}^{4}\left( b+%
\frac{r_{H}}{2}\right) .
\end{equation}%
We observe that the EUP correction term to the volume is negative. This
means that in the presence of the EUP formalism, the volume is always less
than its original form. This result is the opposite of the GUP case, where
the authors showed that correction always increases the volume \cite{13}.

\begin{figure}[tbh]
\begin{minipage}[t]{0.5\textwidth}
        \centering
        \includegraphics[width=\textwidth]{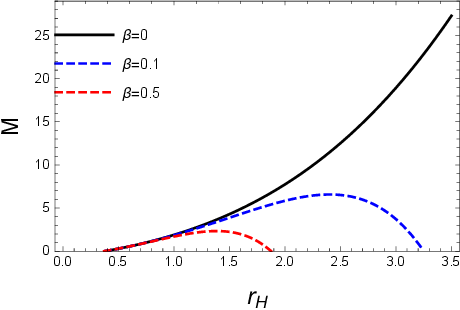}
       \centering{(a) $ a=1$, $b=3/8\pi$.}\label{figEUPMassa}
   \end{minipage}%
\begin{minipage}[t]{0.50\textwidth}
        \centering
        \includegraphics[width=\textwidth]{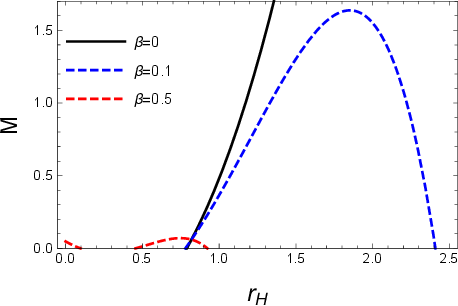}\label{figEUPMassb}\\
        \centering{(b) $ a=1/2\pi$, $b=1$. }
    \end{minipage}\hfill
\caption{The variation of the EUP-VdW black hole mass versus $r_{H}$ for $%
P=0.1.$}
\label{figEUPMass}
\end{figure}

A generalization such as increase or decrease does not apply to the mass
function. To demonstrate our conclusion, in Fig. \eqref{figEUPMass} we plot
the EUP-corrected mass function versus the event horizon with two different
parameter choices that correspond to $a>b$ and $a<b$ cases, respectively. 

\newpage We see that EUP-corrected mass term is limited with an upper-bound
event horizon. However, such a constraint does not exist in the ordinary
case. Furthermore, we observe that the greater deformation parameter
decreases the physical event horizon interval.

\begin{figure}[tbh]
\begin{minipage}[t]{0.5\textwidth}
        \centering
        \includegraphics[width=\textwidth]{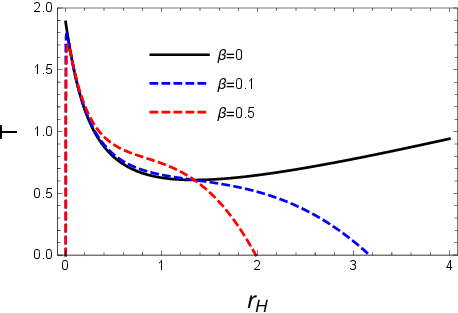}
       \centering{(a) $ a=1$, $b=3/8\pi$.}\label{figEUPTa}
   \end{minipage}%
\begin{minipage}[t]{0.5\textwidth}
        \centering
        \includegraphics[width=\textwidth]{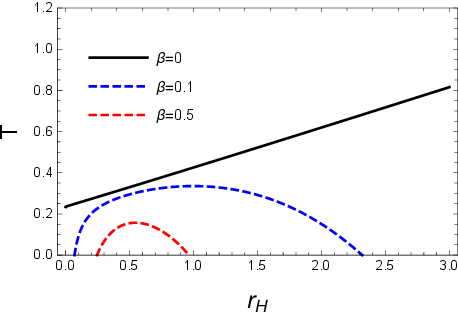}\label{figEUPTb}\\
        \centering{(b) $ a=1/2\pi$, $b=1$. }
    \end{minipage}\hfill
\caption{The variation of the EUP-corrected Hawking temperature versus $%
r_{H} $ for $P=0.1.$}
\label{figEUPT}
\end{figure}

Then, using Eq. (\ref{h}), we get the EUP-corrected Hawking temperature 
\begin{eqnarray}
T &=&\frac{a}{2r_{H}}+2r_{H}P+\beta \frac{ar_{H}}{2}+\frac{ab}{(3b+2r_{H})}%
\left[ \frac{3\beta b}{4}-\frac{2}{r_{H}}-2\beta r_{H}-\frac{81}{32}\frac{%
\beta b^{4}}{r_{H}}\right]  \notag \\
&+&\frac{ab^{2}}{(3b+2r_{H})^{2}}\left[ \frac{81}{16}\beta b^{3}+\frac{3}{2}%
\beta r_{H}+\frac{3}{16r_{H}}\left( 8-45b^{2}\beta \right) \right]  \notag \\
&+&\frac{9}{16}\frac{a\beta b^{2}}{r_{H}}-\frac{b}{2}(a\beta -4P)+\frac{%
a\beta }{4}r_{H}-6\beta r_{H}^{2}P\left( b+\frac{r_{H}}{2}\right) .
\label{TH}
\end{eqnarray}
In the $\beta =0$ limit, it reduces to 
\begin{eqnarray}
T_{HUP} &=&\frac{a}{2r_{H}}+2 P (r_{H}+b)-\frac{2ab}{r_{H}(3b+2r_{H})} +%
\frac{3ab^{2}}{2r_{H}(3b+2r_{H})^{2}}.  \label{THup}
\end{eqnarray}
We demonstrate the Hawking temperature behavior versus event horizon radii
in Fig. \ref{figEUPT}. 

\newpage In the ordinary case, the Hawking temperature takes positive values
for all event horizon radii, whereas this characteristic loses its validity
when the EUP corrections are taken into account. In the new formalism, the
Hawking temperature becomes physical in a certain range. At larger
deformations, the width of this range decreases.

Afterward, we derive the EUP-corrected entropy of the VdW black hole by
employing Bekestein's area law. According to the semi-classical formulation 
\begin{equation}
S=\int \frac{dM}{T},  \label{S}
\end{equation}%
we substitute Eqs. (\ref{TH}) and (\ref{MA}) in Eq. (\ref{S}). We find
results%
\begin{equation}
S=\frac{\pi }{\beta }\log \left( 1+\beta r_{H}^{2}\right).
\end{equation}%
Then, we expand it in the power series of the deformation parameter and
discard the higher terms of order $\mathcal{O}\left( \beta ^{2}\right) $.
This yields: 
\begin{equation}
S\simeq \pi r_{H}^{2}-\frac{1}{2}\pi \beta r_{H}^{4}.
\end{equation}%
We conclude that the correction term of the EUP formalism slows down the
rate of increase of the entropy function. This result was also obtained by
the authors in the GUP formalism \cite{13}.

\begin{figure}[tbh]
\begin{minipage}[t]{0.5\textwidth}
        \centering
        \includegraphics[width=\textwidth]{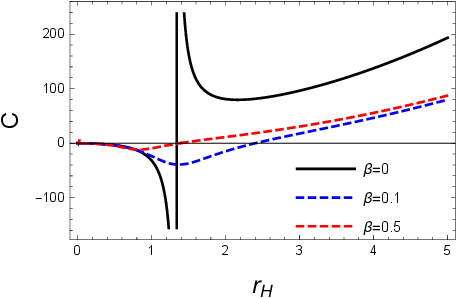}
       \centering{(a) $ a=1$, $b=3/8\pi$ .}\label{figEUPCa}
   \end{minipage}%
\begin{minipage}[t]{0.50\textwidth}
        \centering
        \includegraphics[width=\textwidth]{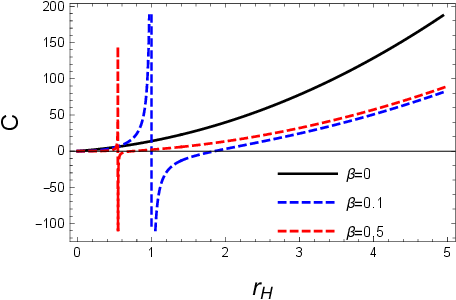}\label{figEUPCb}\\
        \centering{(b) $ a=1/2\pi$, $b=1$ }
    \end{minipage}\hfill
\caption{The variation of the EUP-corrected heat capacity versus $r_{H}$ for 
$P=0.1.$ }
\label{figEUPC}
\end{figure}

The heat capacity is an important thermal function for black hole
thermodynamics because the thermal stability of a black hole is analyzed by
the behavior of its heat capacity. A positive heat capacity indicates the
black hole's stability, while a negative heat capacity indicates its
instability, so the heat capacity gives us clues about a possible phase
transition. We calculate the heat capacity by the following formula 
\begin{equation}
C=\frac{dM}{dT_{H}},
\end{equation}%
and obtain the EUP-corrected heat capacity 
\begin{equation}
C=4\pi \frac{4\pi r_{H}^{2}P-\frac{1}{2}h\left( r_{H},P\right) -\frac{r_{H}}{%
2}\frac{\partial }{\partial r_{H}}h\left( r_{H},P\right) }{\frac{\partial }{%
\partial r_{H}}\left( 8\pi r_{H}P-\frac{1}{r_{H}}h\left( r_{H},P\right) -%
\frac{\partial }{\partial r_{H}}h\left( r_{H},P\right) \right) \left(
1+\beta r_{H}^{2}\right) }.
\end{equation}%
Here, Eq. \eqref{h} has to be used to get the explicit form of this
functional heat capacity, however, it is very complicated. Therefore, we
decide to present it in its current form and analyze it with numerical
methods. To this end, in Fig. \ref{figEUPC}, we display a graphical
representation of the heat capacity for $a>b$ and $b>a$ cases. 

We observe that EUP-corrected VdW black holes are unstable in the $a>b$
case. However, we see stability and instability together in the $b>a$ case.

\section{Limiting cases}

\label{sec4} In this section, we extend our investigation and explore two
sub-scenarios.

\subsection{EUP-Schwarzschild AdS black hole}

\begin{figure}[tbph]
\centering
\includegraphics[scale=1.2]{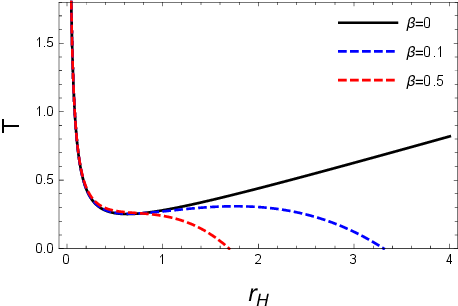}
\caption{The variation of the EUP-corrected Hawking temperature of the AdS
Schwarzschild black hole versus $r_{H}$ for $P=0.1.$}
\label{EUPSCH}
\end{figure}

To start off the first one, we set $b=0$, and consider $a=1/2\pi $ for the
sake of convenience. Such a choice of $b$ does not violate the reverse
isoperimetric inequality. In this scenario, the equation of state given in
Eq. (\ref{eq}), takes the form of 
\begin{equation}
T=\left( P+\frac{a}{v^{2}}\right) v.
\end{equation}%
Then, the lapse function, given in Eq. (\ref{METFUN}), reduces to%
\begin{equation}
f\left( r\right) =1-\frac{2M}{r}+\frac{8\pi }{3}r^{2}P+ \frac{\beta r^2}{6}%
\left(1-24 \pi r^2 P\right).
\end{equation}%
Considering that the last term, which is linearly proportional to beta, is
the EUP-correction term, we can say that this lapse function is the
EUP-corrected AdS Schwarzschild black hole lapse function. Using this lapse
function we express the EUP-corrected mass, volume, and Hawking temperature
functions, respectively as follows: 
\begin{eqnarray}
M&=&\frac{r_{H}}{2}+\frac{4\pi }{3}Pr_{H}^{3}+\frac{\beta r_H^3}{12}%
\left(1-24 \pi P r_H^2 \right), \\
V&=&\frac{4\pi }{3}r_{H}^{3}-2\pi \beta r_{H}^{5}, \\
T&=&\frac{1}{4\pi r_{H}}+\frac{3}{8\pi }\beta r_{H}+2r_{H}\left( 1-\frac{3}{2%
}\beta r_{H}^{2}\right) P.
\end{eqnarray}
In Fig. \ref{EUPSCH}, we demonstrate the EUP-corrected Hawking temperature
of AdS Schwarzschild black hole. 

We observe that EUP corrections lead to a physical finite range for the
event horizon radius. At the highest value of the radius, the black hole has
a finite remnant mass. Since such scenarios are widely investigated in the
literature \cite{Park, HH, HC}, we do not prefer to carry on a detailed
discussion here.

\subsection{The ideal gas case}

\begin{figure}[tbph]
\centering
\includegraphics[scale=1.2]{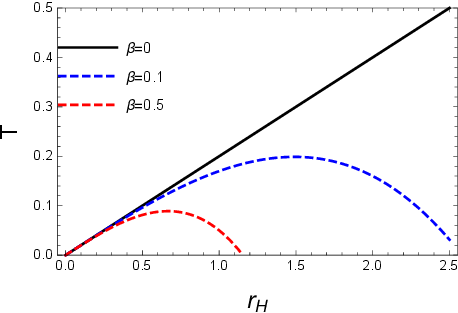}
\caption{The variation of the EUP-corrected Hawking temperature of the AdS
ideal gas black hole versus $r_{H}$ for $P=0.1.$}
\label{IdealHawking}
\end{figure}

In the second scenario, we take $a=0$ and $b=0$, so that Eq. (\ref{eq})
turns to the ideal gas equation of state 
\begin{equation}
T=Pv.
\end{equation}%
In this case, the EUP-corrected lapse function reads 
\begin{equation}
f\left( r\right) =\frac{8\pi }{3}r^{2}P-\frac{2M}{r}-4\pi \beta r^{4}P.
\end{equation}%
Therefore, we can express the EUP-corrected mass, volume, and temperature
function of the ideal gas black hole in the AdS space as follows: 
\begin{eqnarray}
M &=&\frac{4\pi }{3}Pr_{H}^{3}-2\pi \beta Pr_{H}^{5}, \\
V &=&\frac{4\pi }{3}r_{H}^{3}-2\pi \beta r_{H}^{5}, \\
T &=&2r_{H}P\left( 1-\frac{3}{2}\beta r_{H}^{2}\right) .
\end{eqnarray}%
We find that the last terms of all given functions correspond to the EUP
corrections. For the demonstration, we depict the Hawking temperature in
Fig. \ref{IdealHawking}. 

\newpage In the ordinary ideal gas black hole, we observe a linear increase
in the Hawking temperature. However, within the EUP formalism, the
correction terms reduce the Hawking temperature with a term proportional to
the cube of the event horizon. Also, as with the other two black holes, the
temperature can take values in a range where the event horizon remains
physically meaningful. When the event horizon rises to the physical limit, a
remnant mass is left from the black hole.

\subsection{\textcolor{red}{EUP-Schwarzschild black hole}}

\textcolor{red}{It is worth noting that for $a=1/2\pi $, $b=0$ and $l\rightarrow \infty $, the derived lapse function given in Eq. \eqref{METFUN}, converts to the EUP-Schwarzschild case lapse function. \begin{equation}
f\left( r\right) =1-\frac{2M}{r}+\frac{\beta r^{2}}{6}.
\end{equation}
In this case, the mass and temperature of the EUP-Schwarzschild black hole is represented by the following equations: 
\begin{eqnarray}
M &=&\frac{r_{H}}{2}+\frac{\beta r_{H}^{3}}{12}, \\
T &=&\frac{1}{4\pi r_{H}}+\frac{3\beta }{8\pi }r_{H}.
\end{eqnarray}
} 
\textcolor{red}{For a complete analysis, we present the variation of the Hawking temperature of the EUP-Schwarzschild black hole in Fig. \ref{SCH}. We observe that the EUP-corrected temperature is divergent at the highest value of the horizon radius.
\begin{figure}[tbph]
\centering
\includegraphics[scale=1.2]{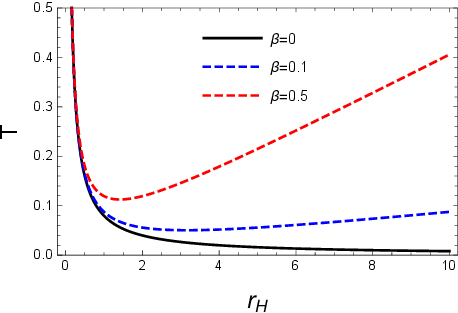}
\caption{The variation of the EUP-corrected Hawking temperature of the
Schwarzschild black hole versus $r_{H}$.}
\label{SCH}
\end{figure}}

\newpage

\section{Conclusion}

The negative cosmological constant can be thought of as thermodynamic
pressure in the context of extended phase space. This makes it possible to
define a black hole equation of state and compare it with a fluid or gas
equation of state. In the last decade, authors have derived asymptotically
AdS black hole metrics whose thermal quantities precisely mimic that of a
VdW fluid. Some other authors handled the VdW black hole thermodynamics with
a quantum mechanical deformation, namely the generalized uncertainty
principle, which predicts a minimal measurable length. They concluded that
the considered quantum deformation plays an important role in the
thermodynamics of the VdW black hole.

In this manuscript, we examine the VdW black hole with the EUP formalism,
which allows a minimal momentum uncertainty value. Our results show that
this kind of quantum deformation sets an upper event horizon limit value,
unlike what exists in the literature. Based on this result, the Hawking
temperature and the black hole mass become physically meaningful within a
certain event horizon radius range. EUP formalism also affects entropy by
slowing down the rate of its increase. Depending on the other parameters VdW
black hole can be unstable or stable. Furthermore, our study also shed light
on other sub-scenarios such as AdS Schwarzschild, ideal gas black holes,
EUP-Schwarzschild and their thermodynamics.

\section*{Acknowledgments}

This work is supported by the Ministry of Higher Education and Scientific
Research, Algeria under the code: B00L02UN040120230003. B. C. L\"{u}tf\"{u}o%
\u{g}lu is grateful to the P\v{r}F UHK Excellence project of 2211/2023-2024
for the financial support.

\section*{Data Availability Statements}

The authors declare that the data supporting the findings of this study are
available within the article.

\section*{Competing interests}

The authors declare no competing interests.


\bigskip

\begin{thebibliography}{99}
\bibitem{amati} D. Amati, M. Ciafaloni, G. Veneziano, Phys. Lett. B \textbf{%
216}, 41 (1989).

\bibitem{roveli} C. Rovelli, Living Rev. Rel. \textbf{1}, 1 (1998).

\bibitem{Girelli} F. Girelli, E. R. Livine, D. Oriti, Nucl. Phys. B \textbf{%
708}, 411 (2005).

\bibitem{kempf} A. Kempf, J. Math. Phys. \textbf{35}, 4483 (1994).

\bibitem{kempf1} H. Hinrichsen, A. Kempf, J. Math. Phys. \textbf{37}, 2121
(1996).

\bibitem{kempf2} A. Kempf, J. Math. Phys. \textbf{38}, 1347 (1997).


\bibitem{Bolen} B. Bolen, M. Cavaglia, Gen. Relativ. Gravit. \textbf{37},
1255 (2005).

\bibitem{Park} M. -I. Park, Phys. Lett. B \textbf{659}, 698 (2008).


\bibitem{S. Mignemi} S. Mignemi, Mod. Phys. Lett. A \textbf{25}, 1697 (2010).

\bibitem{1} S. W. Hawking, Common. Math. Phys. \textbf{43}, 199 (1975).

\bibitem{2} S. W. Hawking, Phys. Rev. D \textbf{13}, 191 (1976).

\bibitem{3} J. B. Hartle, S. W. Hawking, Phys. Rev. D \textbf{13}, 2188
(1976).

\bibitem{can1} A. D. Helfer, Phys. Rev. D \textbf{100}, 025005 (2019).

\bibitem{can2} N. Farahani, H. Hassanabadi, W. S. Chung, B. C. L\"{u}tf\"{u}o%
\u{g}lu, S. Zarrinkamar, EPL \textbf{132}, 50009 (2020).

\bibitem{can3} J. Pinochet, Phys. Educ. \textbf{56}, 053001 (2021).

\bibitem{can4} W. X. Chen, J. X. Li, J. Y. Zhang, Int. J. Theor. Phys. 
\textbf{62}, 96 (2023).

\bibitem{4} J. D. Bekenstein, Phys. Rev. D \textbf{7}, 2333 (1973).

\bibitem{5} J. D. Bekenstein, Phys. Rev. D \textbf{9}, 3292 (1974).


\bibitem{6} S. W. Hawking, D. N. Page, Commun. Math. Phys. \textbf{87}, 577
(1983).

\bibitem{7} D. Kubiznak, R. B. Mann, J. High Energy Phys. \textbf{1207}, 033
(2012).

\bibitem{8} S. Gunasekaran, R. B. Mann, D. Kubiznak, J. High Energy Phys. 
\textbf{1211}, 110 (2012).

\bibitem{9} A. Rajagopal, D. Kubiznak, R. B. Mann, Phys. Lett. B \textbf{737}%
, 277 (2014).

\bibitem{10} T. Delsate, R. B. Mann, J. High Energy Phys. \textbf{02}, 070
(2015).

\bibitem{Parthapratim} P. Pradhan, EPL \textbf{116}, 10001 (2016).

\bibitem{Huchen} Y. Hu, J. Chen, Y. Wang, Gen. Relativ. Gravit. \textbf{49},
148 (2017).

{\color{red} }

\bibitem{Luciano} G. G. Luciano, A. Sheykhi, arXiv:2304.11006 [hep-th].

\bibitem{mangal} N. Altamirano, D. Kubiznak, R. B. Mann, Z. Sherkatghanad,
Galaxies \textbf{2}, 89 (2014).

\bibitem{11} M. R. Setare, H. Adami, Phys. Rev. D \textbf{91}, 084014 (2015).

\bibitem{12} U. Debnath, Eur. Phys. J. Plus \textbf{135}, 424 (2020).

\bibitem{13} \"O. \"Okc\"u, E. Aydiner, Int. J. Theor. Phys. \textbf{59},
2839 (2020).

\bibitem{Raimundo} R. N. Costa Filho, J. P. M. Braga, J. H. S. Lira, J. S.
Andrade Jr., Phys. Lett. B \textbf{755}, 367 (2016).

\bibitem{bilel} B. Hamil, M. Merad, Eur. Phys. J. Plus \textbf{133}, 174
(2018).

\bibitem{bilel3} B. Hamil, M. Merad, Int. J. Mod. Phys. A \textbf{33},
1850177 (2018).

\bibitem{bilel1} B. Hamil, M. Merad, T. Birkandan, Eur. Phys. J. Plus 
\textbf{134}, 278 (2019).

\bibitem{Mur} J. R. Mureika, Phys. Lett. B \textbf{789}, 88 (2019).

\bibitem{bilel2} B. Hamil, M. Merad, Few-Body Syst. \textbf{60}, 36 (2019).

\bibitem{Jaume} J. Gin\'{e}, G. G. Luciano, Eur. Phys. J. C \textbf{80},
1039 (2020).

\bibitem{Mariusz} M. P. Dabrowski, F. Wagner, Eur. Phys. J. C \textbf{80},
676 (2020).

\bibitem{Ali} Y. Kumaran, A. \"Ovg\"un, Chin. Phys. C \textbf{44}, 025101
(2020).

\bibitem{Hassan1} B Hamil, B. C. L\"{u}tf\"{u}o\u{g}lu, H. Aounallah, Mod.
Phys. Lett. A \textbf{36}, 2150021 (2021).

\bibitem{Hassan2} H. Hassanabadi, W. S. Chung, B. C. L\"{u}tf\"{u}o\u{g}lu,
E. Maghsoodi, Int. J. Mod. Phys. A \textbf{36}, 2150036 (2021).

\bibitem{Ozgur} \"O. \"Okc\"u, E. Aydiner, EPL \textbf{138}, 39002 (2022).

\bibitem{Hassan} H. Chen, H. Hassanabadi, B. C. L\"{u}tf\"{u}o\u{g}lu, Z.
-W. Long, Gen. Relativ. Gravit. \textbf{54}, 143 (2022).

\bibitem{dahbi} B Hamil, B. C. L\"{u}tf\"{u}o\u{g}lu, L. Dahbi, Int. J. Mod.
Phys. A \textbf{37}, 2250130 (2022).

\bibitem{Pet1} L. Petruzziello, Phys. Lett. B \textbf{883}, 137293 (2022).

\bibitem{HH} E. Maghsoodi, H. Hassanabadi, W. S. Chung, EPL \textbf{129},
59001 (2020).

\bibitem{HC} B. Hamil, B. C. L\"{u}tf\"{u}o\u{g}lu, EPL \textbf{134}, 50007
(2021).

\bibitem{Xiang} L. Xiang, X. Q. Wen, J. High Energy Phys. \textbf{10}, 046
(2009).
\end{thebibliography}
\end{document}